\documentclass{article}%
\usepackage{amsmath}
\usepackage{amsfonts}
\usepackage{amssymb}
\usepackage{graphicx}%
\providecommand{\U}[1]{\protect\rule{.1in}{.1in}}

\begin{document}

\title{Quantum Nonlocality and Indistinguishability}
\author{Luiz Carlos Ryff\\\textit{Instituto de F\'{\i}sica, Universidade Federal do Rio de Janeiro,}\\\textit{Caixa Postal 68528, 21041-972 Rio de Janeiro, Brazil}\\E-mail: ryff@if.ufrj.br}
\maketitle

\begin{abstract}
It is shown how a \textquotedblleft meddlesome\textquotedblright\ photon
indistinguishable from another photon of an entangled pair can affect the
result of an Einstein-Podolsky-Rosen (EPR) experiment. This makes it clear the
importance of the notion of field over that of particle.


\end{abstract}

Quantum nonlocality, in which acting on a particle of an entangled pair we can
\textquotedblleft force\textquotedblright\ \textrm{[1]} the other into a
well-defined state, is one of the amazing consequences of quantum formalism.
This can be accomplished via EPR correlations \textrm{[2]}. Although it has
been confirmed by experiment, its interpretation is still a matter of dispute.
For Bell \textrm{[3] }and Bohm \textrm{[4]} there should be some kind of
interaction between the particles, but not all physicists share the same point
of view \textrm{[5]}. Experiments have been performed to try to determine a
lower limit to the speed of this possible interaction \textrm{[6]}, and it has
been demonstrated that if this speed is finite then superluminal signaling
would be possible, at least in principle \textrm{[7]}. Quantum teleportation
\textrm{[8] }and entanglement swapping \textrm{[9]} are important offsprings
of quantum nonlocality. Actually, as has been shown, in certain circumstances
the very same phenomenon can be seen as quantum teleportation, as entanglement
swapping, or as usual EPR correlation, depending on the Lorentz frame from
which it is observed \textrm{[10]}.

Indistinguishability plays a crucial role in quantum mechanics. In classical
physics two particles, even being identical, have independent identities,
which is reflected in Maxwell--Boltzmann statistics. On the other hand, this
is not true in quantum mechanics, which leads to Bose-Einstein and Fermi-Dirac
statistics. In fact, the very concept of particle is somewhat diffuse in this
case. Lamb duly criticized the idea of photon \textrm{[11]}, and we have to be
careful not to say (as occurs with some frequency) that a \textit{quantum
particle} (which may be a photon, an electron, an atom, and even a
molecule)\ can be at two different places at the same time (as in a two-slit
experiment, for instance). To avoid some apparent paradoxical conclusions
\textrm{[12]} it would be advisable to keep Ketterle teaching in mind: we
prepare waves and detect particles \textrm{[13]}. I intend to discuss a
consequence of the mathematical formalism of quantum mechanics\ here, which
involves quantum nonlocality and indistinguishability, that corroborates this
point of view. Although it is a simple result, to my knowledge it has not been
previously discussed in the literature and may have important consequences for
questions related to interpretational matters, and possibly to the field of
quantum communication as well.%
\[%
{\parbox[b]{4.7193in}{\begin{center}
\includegraphics[
height=2.7415in,
width=4.7193in
]%
{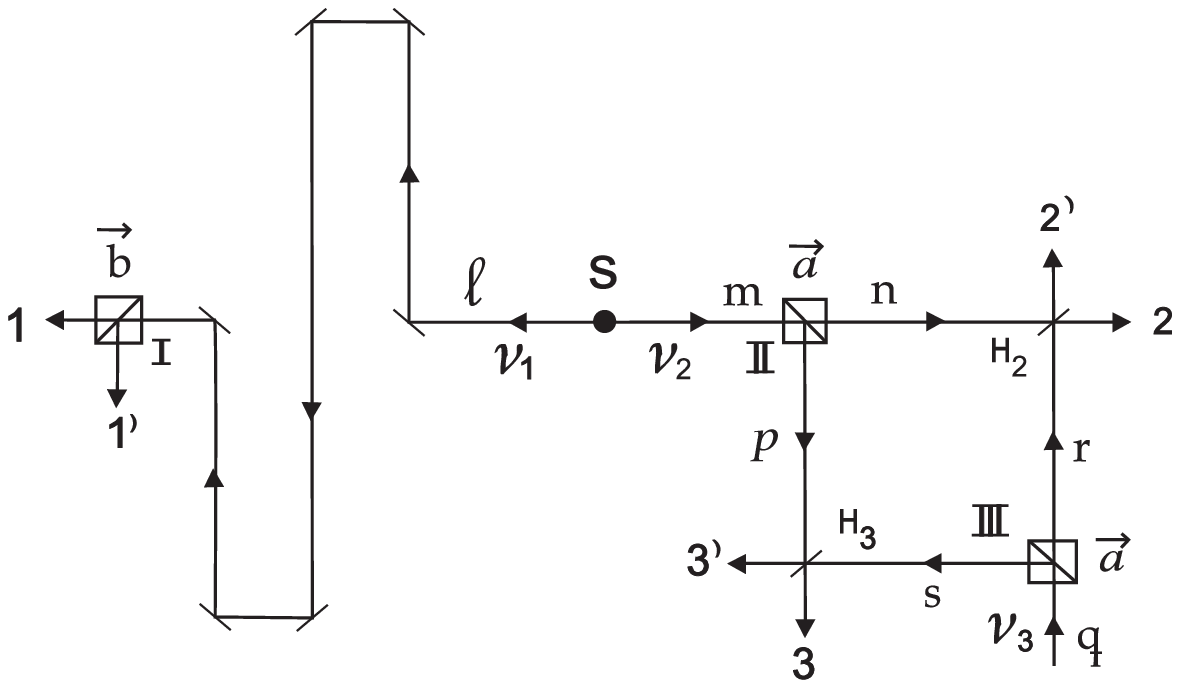}%
\\
Fig.1
\end{center}}}
\]

Let us consider the experiment represented in Fig.1. A source $S$ emits pairs
of photons in the polarization-entangled state%
\begin{equation}
\mid\varphi\rangle=\frac{1}{\sqrt{2}}(\mid a_{\parallel}\rangle_{1}\mid
a_{\parallel}\rangle_{2}+\mid a_{\perp}\rangle_{1}\left\vert a_{\perp
}\right\rangle _{2}),\tag{1}%
\end{equation}
where $\mid a_{\parallel}\rangle_{1}$ ($\mid a_{\perp}\rangle_{1}$) represents
a photon $\nu_{1}$ with linear polarization $\mathbf{a}$ ($\mathbf{a}_{\perp}%
$) (and so on). Photon $\nu_{1}$ ($\nu_{2}$) impinges on polarizer $I$ ($II$)
, oriented parallel to $\mathbf{b}\ $($\mathbf{a}$). Photon $\nu_{1}$
\textrm{[14]} follows via a detour, so that the detections of $\nu_{1}$ and
$\nu_{2}$ are time-like events. The importance of considering time-like events
when discussing EPR correlations has previously been stressed \textrm{[15]}.
This allows us to know which measurement \textit{really} forces the other
photon into a well-defined polarization state. (Here the word
\textquotedblleft measurement\textquotedblright\ is an abuse of language,
since the photon has no previous polarization to be measured. Actually, the
measurement forces the photon into a well-defined polarization state.) A
photon $\nu_{3}$ with linear polarization $\mathbf{c}$ (except for its
polarization state, in all respects identical to $\nu_{2}$) impinges on
polarizer \textrm{III}. For argumentation purposes, it is sufficient to
consider the situation in which detections at $\mathbf{2}$ and $\mathbf{3}$
occur at the same time in the laboratory frame \textrm{[16]}. In this case, it
is not possible to know whether $\nu_{2}$ has been transmitted or reflected at
polarizer \textrm{II}. We could be led to infer (erroneously, as we will see)
that $\nu_{1}$ will always impinge on polarizer \textrm{I} either in state
$\mid a_{\parallel}\rangle$ or state $\mid a_{\perp}\rangle$, since $\nu_{2}$
will necessarily be transmitted or reflected. However, according to the
mathematical formalism of quantum mechanics, by playing with the initial
polarization state of $\nu_{3}$, we can force $\nu_{1}$ into different
polarization states. This result may be interpreted as corroborating the
standpoint that ascribes an essential role to information in quantum
mechanics. But this information has to be seen as corresponding to an
objective fact, that can be translated into subjective knowledge. Amazingly,
the formalism of quantum mechanics gives us no hint about how nature deals
with this information: How does nature \textquotedblleft
know\textquotedblright? How is the information conveyed? These are questions
that do not seem to have a simple answer. Let us then see what mathematical
formalism has to say in the present case.

The initial three photon state can be represented as
\begin{equation}
\mid\psi\rangle=N\left[  \left(  \mid a,l\rangle\mid a,m\rangle+\mid a_{\perp
},l\rangle\mid a_{\perp},m\rangle\right)  \mid c,q\rangle+ST\right]  , \tag{2}%
\end{equation}
where $\mid a,l\rangle\mid a,m\rangle$ $\equiv\mid a_{\parallel},l\rangle
_{1}\mid a_{\parallel},m\rangle_{2}$, $\ \mid a_{\perp},l\rangle\mid a_{\perp
},m\rangle$ $\equiv\mid a_{\perp},l\rangle_{1}\mid a_{\perp},m\rangle_{2}$ and
$\mid c,q\rangle\equiv\mid c_{\parallel},q\rangle_{3}$, and $l$, $m$, and $q$
represent the paths followed by the photons. $ST$ stands for \textit{S}%
ymmetric \textit{T}erms, and $N$ is a normalization factor. The time evolution
of the system is given by
\[
\mid\psi\rangle^{\underrightarrow{pol.II,pol.III}}N\left\{  \left(  \mid
a,l\rangle\mid a,n\rangle+\mid a_{\perp},l\rangle\mid a_{\perp},p\rangle
\right)  \left[  \cos\left(  a,c\right)  \mid a,r\rangle\right.  \right.
\]%
\begin{equation}
\left.  \left.  +\sin\left(  a,c\right)  \mid a_{\perp},s\rangle\right]
+ST\right\}  \equiv\mid\psi^{\prime}\rangle\tag{3}%
\end{equation}
and
\[
\mid\psi^{\prime}\rangle^{\underrightarrow{H_{2},H_{3}}}N\left\{  \left(  \mid
a,l\rangle\mid2\rangle+\mid a_{\perp},l\rangle\mid3\rangle\right)  \left[
\cos\left(  a,c\right)  \mid2\rangle\right.  \right.
\]%
\begin{equation}
\left.  \left.  +\sin\left(  a,c\right)  \mid3\rangle\right]  +...+ST\right\}
\equiv\mid\psi^{\prime\prime}\rangle. \tag{4}%
\end{equation}
Here, $\mid2\rangle$ represents a photon following direction $\mathbf{2}$ and
so on, and $H_{2}$ and $H_{3}$ are 50\%:50\% beam-splitters. Hence,
\begin{equation}
\mid\psi^{\prime\prime}\rangle=N\left[  \sin\left(  a,c\right)  \mid
a,l\rangle\mid2\rangle\mid3\rangle+\cos\left(  a,c\right)  \mid a_{\perp
},l\rangle\mid3\rangle\mid2\rangle+...+ST\right]  . \tag{5}%
\end{equation}
Now, taking into account the \textit{ST}, we can write:
\begin{equation}
\mid\psi^{\prime\prime}\rangle=N\left[  \sin\left(  a,c\right)  \mid
a,l\rangle\mid2\rangle\mid3\rangle+\cos\left(  a,c\right)  \mid a_{\perp
},l\rangle\mid2\rangle\mid3\rangle+...\right]  . \tag{6}%
\end{equation}
Therefore, whenever coincident detections occur at $\mathbf{2}$ and
$\mathbf{3}$, $\nu_{1}$ is forced into state
\begin{equation}
\mid\varphi^{\prime}\rangle=\sin\left(  a,c\right)  \mid a,l\rangle
+\cos\left(  a,c\right)  \mid a_{\perp},l\rangle. \tag{7}%
\end{equation}
This is a simple and interesting result (naturally, other coincident
detections can be managed in a similar way). If we remain too closely attached
to the photon picture, $(7)$ may look paradoxical. One may be led to reason as
follows: If $\nu_{2}$ is detected at $\mathbf{2}$ ($\mathbf{3}$) it has been
transmitted (reflected), which would imply, according to $(1)$, that $\nu_{1}$
has been forced into state $\mid a\rangle$ ($\mid a_{\perp}\rangle$). But a
photon has no individuality, and it is not possible to know where $\nu_{2}$
has \textit{really} been detected (actually, this question has no meaning).
The realization of the experiment that has been discussed here would be an
impressive and palpable demonstration of the indistinguishability of photons,
and would corroborate the point of view that emphasizes the importance of the
notion of field over that of particle.


\begin{thebibliography}{99}                                                                                               %


\bibitem[1]{}P. A. M. Dirac, \textit{The Principles of Quantum Mechanics}, 4th
edn. (Oxford University Press, Oxford, 1958).

\bibitem[2]{}J. S. Bell, \textit{Speakable and Unspeakable in Quantum
Mechanics} (Cambridge University Press, Cambridge, 1987).

\bibitem[3]{}J. S. Bell, in \textit{The Ghost in the Atom}, eds. P. C. Davies
and J. R. Brown (Cambridge University Press, Cambridge, 1989).

\bibitem[4]{}D. Bohm, in \textit{The Ghost in the Atom}, eds. P. C. Davies and
J. R. Brown (Cambridge University Press, Cambridge, 1989).

\bibitem[5]{}J. Bricmont, Making Sense of Quantum Mechanics (Springer
International Publishing, Switzerland, 2016)\textrm{, }and references there
in. (A lucid contribution to the debate on the foundations of quantum theory.
It can be read with interest even by those who do not adhere to the Bohmian interpretation.).

\bibitem[6]{}V. Scarani, W. Tittel, H. Zbinden, and N. Gisin, \textit{Phys.
Lett. A }\textbf{276}, 1 (2000); D. Salart, A. Baas, C. Branciard, N. Gisin,
and H. Zbinden, \textit{Nature} \textbf{454}, 861 (2008); J. Yin, et al.,
\textit{Phys. Rev. Lett. }\textbf{110}, 260407 (2013).

\bibitem[7]{}V. Scarani and N. Gisin, \textit{Phys. Lett. A }\textbf{295}, 167
(2002); V. Scarani and N. Gisin, \textit{Braz. J. Phys.} \textbf{35}, 328
(2005); J.-D. Bancal, et al., \textit{Nat. Phys.}\textbf{ 8}, 867 (2012); T.
J. Barnea, et al., \textit{Phys. Rev. A }\textbf{88}, 022123 (2013); L. C.
Ryff, arXiv: 0903.1076 [quant-ph], and arXiv: 1506.07383 [quant-ph].

\bibitem[8]{}C. H. Bennett, G. Brassard, C. Cr\'{e}peau, R. Jozsa, A. Peres,
and W. K. Wootters, \textit{Phys. Rev. Lett. }\textbf{70}, 1895 (1993); D.
Bouwmeester, J.-W. Pan, K. Mattle, E. Eibl, H. Weinfurter, and A. Zeilinger,
\textit{Nature}\textbf{ 390}, 575 (1997).

\bibitem[9]{}M. Zukowski, A. Zeilinger, and H. Weinfurter, \textit{Ann. N. Y.
Acad. Sci. }\textbf{755}, 91 (1995); J.-W. Pan, D. Bouwmeester, H. Weinfurter,
and A. Zeilinger, \textit{Phys. Rev. Lett.} \textbf{80}, 3891 (1998).

\bibitem[10]{}L. C. Ryff, \textit{Workshop on Mysteries, Puzzles and Paradoxes
in Quantum Mechanics, }Gargnano, Garda Lake, Italy (1999); L. C. Ryff,
\textit{Phys. Rev. A }\textbf{60}, 5083 (1999); L. C. Ryff, \textit{J. Mod.
Opt. }\textbf{48}, 905 (2001).

\bibitem[11]{}W. E. Lamb,Jr., \textit{Applied Phys B} \textbf{60}, 77 (1995).
The interesting article by A. Hobson, \textit{Am. J. Phys.} \textbf{81}, 211
(2013), emphasizing the notion of field over that of particle is worth
reading. Actually, the idea of photon can be useful, provided we use it wisely.

\bibitem[12]{}L. C. Ryff, \textit{Found. Phys. Lett}. \textbf{10} (3), 207
(1997); L. C. Ryff, \textit{Quant. Semiclass. Opt}. \textbf{10}, 409 (1998);
L. C. Ryff, in \textit{Causality and Locality in Modern Physics}, G. Hunter,
S. Jeffers, and J-P. Vigier (eds), Kluwer Academic, Dordrecht (1998); L. C.
Ryff, \textit{Phys. Rev. A }\textbf{52}, 2591 (1995), where it is shown that
in some situations the attempt to trace back the behavior of the photon leads
to a contradiction.

\bibitem[13]{}In a popular talk about the Bose-Einstein condensate, given at
the annual meeting of the German Physical Society in Hannover in March 2003,
the Nobel prize winner Wolfgang Ketterle told the public that it is very hard
to understand quantum mechanics, but after several years of physical practice
one gets used to preparing waves and detecting particles. (See, B. Falkenburg,
\textit{Particle Metaphysics}, Springer, Berlin (2007), p.280).

\bibitem[14]{}Here the term \textit{photon} is a useful abuse of language. It
refers to a Fock state of occupation number one. We have to keep in mind that
we are considering a physical entity that propagates as a wave and is detected
as a particle. Actually, this is the big quantum mystery, the reduction of the
state vector (collapse of the wave function), which is connected to quantum nonlocality.

\bibitem[15]{}L. C. Ryff, \textit{Found. Phys.} \textbf{44} (1), 58 (2014);
arXiv: 1506.07383 [quant-ph].

\bibitem[16]{}Strictly speaking, the detections don't need to be simultaneous,
provided it is impossible to know which photon is detected at which place.
Actually, it is always possible to describe the experiment from a Lorentz
moving frame in which the events are no longer simultaneous. Naturally, to
define \textquotedblleft simultaneous\textquotedblright\ we have to take into
account the lengths of the wave packets associated to the photons.
\end{thebibliography}
\end{document}